\documentclass[doublecol]{epl2}

\newcommand{\p}{\partial}
\newcommand{\myskip}[1]{}
\newcommand{\dec}{{\rm dc}}
\newcommand{\eps}{\varepsilon}
\renewcommand{\d}{{\rm d}}
\newcommand{\K}{{\rm K}}
\newcommand{\Rey}{{\rm R_e}}
\newcommand{\cl}{{\rm cl}}
\renewcommand{\ni}{\noindent}
\newcommand{\nn}{\nonumber}

\newcommand{\kg}{{\rm kg}}
\newcommand{\m}{{\rm m}}
\newcommand{\s}{{\rm s}}

\newcommand{\kpc}{{\rm kpc}}

\newcommand{\ac}{{\rm ac}}

\newcommand{\vf}{{\rm vf}}

\newcommand{\LCDM}{$\Lambda${\rm CDM}\,}

\newcommand{\BEQ}{\begin{eqnarray}}
\newcommand{\EEQ}{\end{eqnarray}}
\newcommand{\BEA}{\begin{eqnarray}}
\newcommand{\EEA}{\end{eqnarray}}
\newcommand{\half}{{\frac{1}{2}}}

\begin{document}

\title{Gravitational hydrodynamics of large scale structure formation}

\shorttitle{Hydrodynamics of structure formation in the early
Universe\date{today} }

\author{
Th. M. Nieuwenhuizen$^{1(a)}$, C. H. Gibson$^{2(b)}$,  \and R. E.
Schild$^{3(c)}$} \shortauthor{Nieuwenhuizen, Gibson \& Schild}

\institute{
  \inst{1} Institute for Theoretical Physics,
  Valckenierstraat 65, 1018 XE Amsterdam, The Netherlands\\
\inst{2}
  Mech. and Aerosp. Eng. \& Scripps Institution
  of Oceanogr. Depts., UCSD, 
  La Jolla, CA 92093, USA\\
  \inst{3} Harvard-Smithsonian Center for Astrophysics,
       60 Garden Street, Cambridge, MA 02138, USA
  }

\pacs{98.80.Bp}{Origin and formation of the universe}

\pacs{95.35.+d}{Dark matter}

\pacs{98.20.Jp}{Globular clusters in external galaxies}

\abstract{ The gravitational hydrodynamics of the primordial
plasma with neutrino hot dark matter is considered as a challenge
to the bottom-up cold dark matter paradigm. Viscosity and
turbulence induce a top-down fragmentation scenario before and at
decoupling. The first step is the creation of voids in the plasma,
which expand to $37$ Mpc on the average now. The remaining matter
clumps turn into galaxy clusters. Turbulence produced at expanding
void boundaries causes a linear morphology of 3 kpc fragmenting
protogalaxies along vortex lines. At decoupling galaxies and
proto-globular star clusters arise; the latter constitute the
galactic dark matter halos and consist themselves of earth-mass
H-He planets. Frozen planets are observed in microlensing and
white-dwarf-heated ones in planetary nebulae. The approach
explains the Tully-Fisher and Faber-Jackson relations, and cosmic
microwave 
temperature fluctuations of micro-Kelvins. }

\maketitle

\email{$^{(a)}$t.m.nieuwenhuizen@uva.nl}

\email{$^{(b)}$cgibson@ucsd.edu}

\email{$^{(c)}$rschild@cfa.harvard.edu }

\section{Introduction}

Structure formation in  the Universe starts in the plasma of
protons, electrons, He atoms and neutrinos, that exists up to some
400,000 yr after the Big Bang, the time of decoupling ($dc$) of
photons from matter (last scattering ($L$) or recombination). Then
the plasma transforms to a neutral gas of H and 24\% by weight
$^4$He, with the neutrinos remaining free streaming. As this
occurs at about four thousand degrees Kelvin, a moderate plasma
temperature, we shall seek an explanation in terms of plasma
physics and gravitational hydrodynamics alone. This embodies a
return to the top-down scenario of large scale structure
formation.

Currently it is assumed that cold dark matter (CDM) also exists
and, clustered before decoupling, has set seeds for baryon
condensation. The so-called the concordance or $\Lambda$CDM model
involves also a cosmological constant or dark energy. It describes
a hierarchical bottom-up approach to structure formation, stars
first, then galaxies, clusters, and, finally, voids.

But observations of dense clumps of ancient small stars in old
globular clusters (OGCs) in all galaxies contradict the \LCDM
predictions that star formation should begin only after about 300
million years of dark ages and that the first stars should be
100-1000 $M_{\odot}$ population III superstars. OGCs do not spin
rapidly so they cannot be condensations, and their small stars
imply gentle flows inconsistent with superstars.  Other
difficulties are posed by empty supervoids with size up to 300 Mpc
reported from radio telescope measurements~\cite{coldspot}, dwarf
galaxies with a lot of dark matter\cite{dwarfgalaxies} and a
preferred axis of evil spin direction (AE) that appears at scales
extending to 1.5 Gpc, a tenth of the horizon scale
~\cite{axisevilOliveira}. Nearly every month new observations
arise that pose further challenges to the \LCDM paradigm:
Correlations in galaxy structures~\cite{Disney08}; absence of
baryon acoustic oscillations in
galaxy-galaxy correlations~\cite{SylosBaryshev}; galaxies formed
already when the universe was 4 -- 5 billion years
old~\cite{younggalaxies}; dwarf satellites that swarm our own
galaxy just like its stars~\cite{dwarfs}.

The recent conclusion by one of us that dark matter particles must
have mass of a few eV and probably are 1.5 eV
neutrinos~\cite{Nneutrino09}, means that dark matter is hot (HDM),
urging once more for an explanation of structure formation from
baryons alone, without a cold dark matter trigger.

We shall discuss such baryonic clustering due to a viscous
instability in the plasma, overlooked by the currently popular
linear models of structure formation. CDM is assumed not to exist,
while HDM, though initially important to maintain the homogeneity
of the plasma, has no role in the structure formation. Central in
our discussion will be the huge plasma viscosity $\nu\sim
5\,10^{27}\m^2\s^{-1}$ arising from photons that scatter from free
electrons. This makes the plasma increasingly viscous, while it is
also expanding with space. At some age before the decoupling an
instability creates the first structures, proto-voids and
proto-galaxy-clusters. At decoupling the viscosity drops to hot
gas values $\sim 10^{13}\m^2\s^{-1}$, which creates further
structures at the Jeans scale and at the new, small viscous scale.

The plan of this Letter is to review modern theory of
gravitational hydrodynamical structure formation, to evaluate the
estimates for various fragmentation scales within Friedman
cosmology, and to compare with observations.

\section{Hydrodynamics}
The description of gravitational structure formation starts with
Jeans 1902. He proposes that scales for gravitational condensation
of a uniform fluid of density $\rho$ must be larger than the Jeans
acoustic scale $L_J = V_S/(\rho G)^{1/2}$, where $V_S\sim
c/\sqrt{3}$ is the sound speed of the plasma and
$G=6.67\,10^{-11}$m$^3$/kg$\,$s$^2$ is Newton's constant. As the
plasma Jeans scale is always larger than the horizon scale of
causal connection, this forbids gravitational structure formation
before decoupling. The Jeans criterion reflects a linear
gravitational instability from acoustics, but it neglects the fact
that self gravitational instability of a gas is absolute
~\cite{Gibson00}. All density variations will grow or decrease
unless prevented by the viscous forces, turbulent forces and
diffusion effects.

Conservation of the specific momentum in a fluid is expressed by
the Navier-Stokes equation,

\begin{equation}
\label{NavStokes}
\frac{\partial \vec v }{ \partial t} =
\nabla B + \vec v\! \times \vec \omega
+ \vec F_{viscous} +  \vec F_{other},
\end{equation}
averaged over system control volumes exceeding the momentum
collision length scale. $B$ is the Bernoulli group of mechanical
energy terms $B = p/\rho + \half v^2 + lw$ and the viscous force
is $\vec{F}_{viscous}=\nu_s\nabla^2\vec{v}+(\frac{1}{3}\nu_s
+\nu_b)\nabla\cdot(\nabla\cdot\vec{v})$, with kinematic shear
viscosity $\nu_s=\eta/\rho$ and bulk viscosity $\nu_b=\zeta/\rho$,
while other fluid forces may arise. The inertial-vortex force per
unit mass $ \vec v \times \vec \omega$, with
$\vec\omega=\nabla\!\times\vec v$, produces turbulence if it
dominates the other forces; for example, $ \Rey\equiv|\vec v
\times \vec \omega| / |\vec F_{viscous}|$ is the Reynolds number.
A large viscosity corresponds to a small Reynolds number, with
universal critical value  ${\rm R_e^{\it c}}\sim 25-100$. For
adiabatic flows the ``lost work'' term $lw$ due to frictional
losses is negligible so the enthalpy $p/\rho$ decreases or
increases to compensate for changes in the kinetic energy per unit
mass $\half v^2$.

The turbulence problem is put in a new perspective by Gibson
~\cite{Gibson96}. Universal similarity laws are explained in terms
of the inertial vortex forces $\vec v\! \times \vec \omega$.
From Eq. (\ref{NavStokes}), turbulence is
defined as an eddy-like state of fluid motion where the
inertial-vortex forces of the flow are larger than any other
forces that tend to damp the eddies out ~\cite{Gibson91,GS09}. By
this definition, irrotational flows are non-turbulent. All
turbulence then cascades from small scales to large because
vorticity is produced at small scales and adjacent eddies with the
same spin induce inertial vortex forces that cause the eddies to
merge and form bigger structures.
Thus,
turbulent motions and energy always cascade from small scales to
large, contrary to standard turbulence theories that include in
turbulence also irrotational flows and motions dominated by other
forces.

\section{Gravitohydrodynamics (GHD)}
In his approach with hydrodynamic and diffusive modelling, Gibson
1996 ~\cite{Gibson96} derives several gravitational Schwarz length
scales for structure formation by Kolmogorian dimensional
analysis. With $\tau_g=1/\sqrt{\rho G}$ the gravitational free
fall time in the Jeans length $L_J=V_S\tau_g$, there occur first
the viscous length $L_{SV}=\tau_g\sqrt{\gamma\nu}$, where $\nu$
the kinematic viscosity and $\gamma$ the rate of the strain, i.
e., the magnitude of $e_{ij}=\half(\p v_i/\p x_j+\p v_j/\p x_i)$.
Second, there is the turbulent length
$L_{ST}=(\eps\tau_g^3)^{1/2}$, where $\eps$ is the rate of energy
dissipation per unit mass, and third, the diffusive length
$L_{SD}=\sqrt{D\tau_g}$, where $D$ is the diffusion coefficient.
Within the acoustic horizon scale, $d_H^\ac\sim ct$ structures can
form at scale $L$ if $d_H^\ac \ge L \ge {\rm
max}(L_{SV},L_{ST},L_{SD})$.

We shall evaluate these scales within the flat Friedman metric
$\d s^2=c^2\d t^2-a^2(t)\d{\bf r}^2$. The Friedman
equation for baryonic and neutrino matter reads

\BEQ \frac{\dot a^2}{H_0^2a^2}\equiv\Omega(a)
=\Omega_\Lambda+\frac{\Omega_B+\Omega_\nu}{a^3}
+\frac{\Omega_\gamma}{a^4}.\qquad
\EEQ

\ni where $\Omega_\nu$ becomes $a$-dependent beyond the Compton
temperature $\sim17,000$ K. With $\d t=\d a/H_0a\sqrt{\Omega(a)}$
the age is $\int_0^t \d t$, while the `angular' distance from us
to an object at redshift $z=1/a-1$ reads $d_A(z)=[c/(1+z)]$
$\int_t^{t_0}\d t/a$.

\ni We adopt the Hubble parameter $h=H_0/[100{\rm km/s\,Mpc}]$
=0.744 favored in Ref.~\cite{Nneutrino09}, so that
$m_\nu=2^{3/4}(G_F)^{1/2}m_e^2= 1.4998$ eV and
$\Omega_\nu=0.111/$$h^{3/2}$$=0.173$. For baryons we take
$\Omega_B=0.02265/h^2=0.0409$ from WMAP5~\cite{WMAP5}, while for
photons $\Omega_\gamma=2.47\cdot10^{-5}/h^{2}=4.46\cdot10^{-5}$.
Finally,
$\Omega_\Lambda=1-\Omega_\nu-\Omega_B-\Omega_\gamma=0.786$ assures
a flat space.

\section{Viscous instability in the plasma}
GHD starts with acknowledging the importance of the photon
viscosity. Because it strongly increases in time,  already before
decoupling the plasma becomes too viscous to follow the expansion
of space~\cite{Gibson96}. Thus a gravitational instability occurs,
that tears the plasma apart at density minima, thus creating
voids. Cosmic (super)voids surround us at any distance and the
furthest observable ones are located at the decoupling redshift.
Presently, voids have a 20 times under-density with respect to the
critical density. In between voids the galaxy clusters are located
on ``pancakes'' that join in superclusters.

The shear viscosity $\nu\equiv\nu_s$ reads for $k_BT\ll
m_ec^2$~\cite{deGrootvLvW}

\BEQ \nu=\frac{\eta}{\rho_B}=\frac{1}{\rho_B}\,
\frac{5m_e^2\zeta(3)(k_BT)^4}{9\pi^3\hbar^5c^2\alpha_{\rm
em}^2n_e},  \EEQ

\ni with $m_{e}$ the electron mass, $\alpha_{\rm em}=1/137$ the
fine structure constant and $n_e=0.76\rho_B/m_N$ the electron
density.
With $n_e\sim\rho_B\sim T^3$, $\nu$ increases as $1/T^2$. At WMAP5
values for decoupling it reaches the huge value
$5.85\,10^{27}\m^2/\s$, while the bulk viscosity $\sim
10^{14}\m^2/\s$ is much smaller. In the plasma the acoustic speed
 $V_S=c/\sqrt{3(1+R_z)}$ with $R_z
={3\Omega_B}/{4\Omega_\gamma(1+z)}$~\cite{WeinbergCosmo}
sets the acoustic horizon scale

\BEQ d_H^{ac}(z)=\frac{1}{1+z}\int_0^{t(z)}\d
t'\frac{V_S(t')}{a(t')}.
\EEQ

\ni At WMAP5 decoupling it takes the value $d_H^{ac}=128\,\kpc$
but, estimating $\gamma=V_S/d_H^{ac}$, the viscous length
$L_{SV}=(\nu V_s/G\rho_Bd_H^{ ac})^{1/2}$ is then
only $76\,\kpc$, 
showing that an instability has occurred. This causes an often
overlooked baryonic structure formation in the
plasma~\cite{Gibson96}. The crossover of $L_{SV}$ and $d_H^{ac}$
occurs when $d_H^{ac}= (V_s\nu/G\rho_B)^{1/3}$. This happens at
$z_\vf=5120$, where $d_H^{ac}=7.3$ kpc is the initial void scale.
It expands by a factor $1+z_\vf$ to become 37 Mpc now, a typical
void size, smaller than the supervoids of 50--300 Mpc observed by
radio telescopes. \LCDM models predict such voids formed last and
full of debris, rather than first and empty as
observed~\cite{coldspot}. Foreground voids will play a role in the
cosmic microwave background (CMB) structure at large angles
~\cite{Cover}, especially due to their neutrino depletion at
$z\sim28$ ~\cite{Nneutrino09}. Voids occur next to condensations
with baryonic clustering mass

\BEQ M_\cl=\frac{\pi}{6}\rho_B (d_H^\ac)^3= \frac{\pi V_s
\nu}{6G}=1.7 \cdot 10^{14}M_\odot,\EEQ

\ni  which corresponds to the baryonic mass of fat {\it galaxy
clusters} (cl). The Reynolds number becomes

\BEQ \label{Reynolds}\Rey =\frac{d_H^{ac}V_S}{\nu}
=\frac{9\pi^3\hbar^5c^2\alpha_{\rm em}^2}{5m_e^2\zeta(3)}\,
\frac{d_H^{ac}V_sn_e\rho_B}{(k_BT)^4}\, \EEQ

\ni At $z_{\vf}$ it equals 158, somewhat above critical. At the
boundaries of the clumps it is much smaller, $\Rey(r,z) =\Rey(z)
\left[{\rho_B(r,z)}/{\rho_{B}(z)}\right]^2$, where $z$ codes the
time, $r$ the local position and the uniform terms refer to the
would-be uniform state. While $\Rey=158$  at $z_\vf$ is already
not large, fragmentation leads to small values at the boundaries,
which enhances the effect. As the voids expand, baroclinic torques
at their boundaries produce vorticity and turbulence due to
misalignment of pressure gradients and density gradients. Pressure
gradients will be normal to void boundaries, but density gradients
need not. The rate of vorticity and turbulence production at the
expanding protosupercluster boundaries is ${\partial
\vec{\omega} / \partial t} =$ $ {\nabla \rho \times \nabla p }/{
\rho^2}$. Observations of the Hubble Ultra Deep Field
~\cite{Elm05} show chains of protogalaxies and spiral clump
clusters, as well as DM filaments, formed in this way.

A connection to turbulence was established ~\cite{BersSreen02} in
a study of the CMB temperature difference between two points at
angular separation $r$, viz. $\langle |\Delta T|^p\rangle\sim
r^{\zeta_p}$, where the average is taken over angles between
0.9$^\circ$ and 4$^\circ$. For $0.1<p\le3$ the exponent reads
$\zeta_p\approx p/3$, as in turbulence. A test of Gaussianity in
CMB, $\langle |\Delta T|^p\rangle\sim  \langle |\Delta
T|^3\rangle^{\zeta_p}$ reveals a marked deviation from the
Gaussian value $\zeta_p=p/3$ in the interval $3<p<12$, with
$\zeta_{12}\approx 2.8$, and coinciding with the $\zeta_p$ of
turbulence~\cite{BersSreen03}.  In ref. ~\cite{Bers06} it is
deduced that the data for the first CMB peak involve
$\Rey\sim100$, in striking agreement with our estimate $\sim 158$.

The pancaked structure of matter in between large voids arises
dynamically since voids expand more than matter.

\section {The Axis  of Evil}
 WMAP data shown alignment of low order multipoles $(\ell=2,3)$
 of the CMB spectrum~\cite{axisevilOliveira,axisevil};
further coincidences occur up to $\ell=17$. In our GHD
approach the AE reflects density gradients of big bang turbulence
~\cite{Gibson05} and mixing~\cite{Gibson04} subject to
universal similarity laws of fossil turbulence and turbulent
mixing ~\cite{Gibson91}.
Such turbulence fossils appear in CMB spherical harmonics axes and
the spin axis of local galaxies clusters.

\section{Towards decoupling}

In the period near last scattering, Helium is already formed, so
the density of protons plus H-atoms is
$n=0.76\,\rho_B/m_N$. The fractional ionization $X=n_e/n$ evolves
according to Eq. (2.3.27) of ~\cite{WeinbergCosmo},

\BEQ \label{dXdT=} \frac{\d X}{\d T}
&=&\frac{n\alpha}{HT}\,\frac{X^2-(1-X)/S}{1+A}+
\left(3-\frac{T}{n}\,\frac{\d n}{\d T}\right)\frac{X}{T}, \\
S&=&n\lambda_{\rm T}^3e^{157,894\,K/T},\qquad
 \lambda_T=\hbar\sqrt{\frac{2\pi}{m_ek_BT}},\nn\\
\alpha&=&\frac{1.4377\,10^{-16}(T/K)^{-0.6166}}
{1+5.085\,10^{-3}(T/K)^{0.5300}}\,\frac{{\rm m}^3}{{\rm s}},\\
A&=&\frac{\alpha\lambda_T^{-3}e^{-39,474\,K/T}} {\Gamma_{2s}+8\pi
H/[\lambda_\alpha^3n(1-X)]}.\nn \EEA

\ni Here $S$ is the Saha function and $\lambda_T$ the thermal
length, while $\alpha$ and $A$ are factors involving
$\Gamma_{2s}=8.22458\,s^{-1}$ the two-photon decay rate of the
H$_{2s}$ level and $\lambda_\alpha=1215$\AA \, the Ly$_\alpha$
wavelength. We added the last term in (\ref{dXdT=}) in order to
allow that $n\neq$ const.$T^3$. Baryonic matter will expand less
after clusters have formed. Let us take the geometric mean between
no and full expansion, thus assuming that the matter clumps expand
up till last scattering at $z_L$ by a factor $\sqrt{a/a_\vf}\le
2.2$, implying $\rho_B=(a_\vf/a)^{3/2}\rho_B^\vf$.

Initially $S\ll1$, so the Saha law $X=1-SX^2$ continues to hold.
H-formation makes $X$ decrease appreciably, from where on we have
to solve Eq. (\ref{dXdT=}). The condition for maximal probability
of last scattering~\cite{WeinbergCosmo} can be formulated as $\d
J/d T=J^2$, where $J=c\sigma_TnX/HT$ involves the Thomson cross
section
$\sigma_T=6\pi(\hbar/\alpha_{em}m_ec)^2=6.6525\,10^{-29}\m^2$.
This fixes the surface of last scattering at $T_L=2862$ K,
$z_L=1050$, compared to $z_{dc}=1090$ from WMAP5, and taking
place at age $t_L=408,000$ yr.
The clump size $L_\cl=d_H^{ac}(z_\vf)\sqrt{a_L/a_\vf}$ corresponds
to an angle $\theta_\cl =180^\circ L_\cl/\pi d_A(z_L)=0.84^\circ$,
or spherical index $\ell_\cl =$ $180^\circ/\theta_\cl =215$, which
agrees with the first CMB peak.
At this moment $X=0.01$ makes the Reynolds number as low as 0.12,
thus exhibiting turbulence throughout the bulk, and predicting
more CMB turbulence at smaller scales.

\section{Magnitude of CMB temperature fluctuations}
The smallness of CMB fluctuations, $\delta T/T\sim10^{-7}$ is one
of the mysteries of cosmology. Indeed, how can it be consistent
with a mass contrast of almost 100\% between clumps and voids?
Presently it is described by inflation, where its size is adjusted
in the initial fluctuation spectrum~\cite{WeinbergCosmo}. In order
to explain it from a physical mechanism, let us notice that not
all clump energy can associate with temperature fluctuation, since
in empty space the temperature already decays with the redshift,
$T(z)=(z+1)T_0$. Compared to voids, extra energy of a clump that
is available for photons must stem from the kinetic energy of the
protons, electrons and He atoms. Their densities are low, at
decoupling $\sim (1+z_\dec)^3\Omega_B\rho_c/m_N\sim300$/cc with
$\rho_c=1.04\cdot10^{-26}\kg/\m^3$ the critical density. At a
temperature $T<T_\vf$ this amounts to an excess kinetic energy of
$\frac{3}{2}k_B(T_\vf-T)$ for each of them, which corresponds to
an energy density
$2.37\,\Omega_B(\rho_{c}/m_N)k_BT_0(1+z_\vf^3)(z_\vf-z)$. Let us
assume that due to scatterings this gets redistributed to the
local photons, thereby causing a perturbation in the photon energy
density at last scattering \footnote{
We neglect here that H atoms have formed in mean time.}, $\delta
u_\gamma(z_L)= 4\Omega_\gamma\rho_cc^2(z_L+1)^4(\delta T/T)_L$.
This yields \BEQ \left.\frac{\delta T}{T}\right|_L=0.593\,
\frac{\Omega_B}{\Omega_\gamma}\frac{k_BT_0}{m_Nc^2}
\frac{(z_\vf+1)^3(z_\vf-z_L)}{(z_L+1)^4}. \EEQ

\ni Though $k_BT_0/m_Nc^2=2.5\cdot 10^{-13}$ is very small, we
find for $z_\vf=5120$, $z_L=1050$ that $\left.\delta
T/T\right|_L=6.1\cdot10^{-8}$, which corresponds presently to
$\delta T_\cl\equiv T_\cl-T_{void}=+0.17\,\mu$K. This is the right
order of magnitude, since $\delta T_\cl ^2=204 \cdot[2\pi/\ell_\cl
(\ell_\cl +1)]\,\mu\K^2$ compares to the first peak of the
correlator, $C_{220}=5800 \cdot[2\pi/\ell_\cl (\ell_\cl
+1)]\,\mu\K^2$ from WMAP5~\cite{WMAP5,Cover}. Voids do not have
this baryonic content, which explains the observed connection hot
spots -- (super)clusters, cold spots -- voids.


\section{Fragmentation in the gas at two scales}
At last scattering, the plasma turns into a neutral gas and further
baryonic structures form. The free fall time is $\tau_g =
 1.68$ Myr, while the age is $t_L =0.41$ Myr. The sound speed of a
monoatomic gas is $V_s=\sqrt{5 p/3\rho}$. For H with 24\% weight
of He, $p=0.82\rho k_BT/m_N$ yields $V_S=5.68$ km/s. The gas
fragments at the Jeans scale $L_J=V_s\tau_{g}=9.78$ pc into
Proto-Globular Clusters (PGCs) with Jeans mass

\BEQ M_{PGC}=\frac{\pi}{6}\rho_{B}L_J^3= \frac{\pi
V_s^3}{6G^{3/2}\rho_B^{1/2}}= 
38,000\,M_\odot.\EEQ

\ni  This Jeans cluster formation is well known, but
not always welcomed. In our approach it is a standard fragmentation.

At decoupling the viscosity decreases from photon viscosity values
to hot-gas values. The He viscosity can be estimated as $\eta_{\rm
He}(T_L)=5.9\cdot10^{-5}{\kg}/{\m\s}$. For the 76:24 H-He mixture
it will be about $0.76/8+0.24/4=0.155$ of this.
The critical viscous length
$L_{SV}=(V_s\eta/G\rho_B^2)^{1/3}=3.9\cdot 10^{14}$ m implies a
condensation mass

\BEQ M=\frac{\pi}{6}\rho_B L_{SV}^3=\frac{\pi
V_s\eta}{6G\rho_B}=13M_\oplus=3.9\cdot10^{-5}M_\odot.\EEQ

\ni We may call these objects H-He planets, Primordial Fog
Particles or Milli Brown Dwarfs. Their mass is in good agreement
with estimates from microlensing of a distant quasar
~\cite{Schild96,Schild06} and so-called cometary knots in the
Helix nebula ~\cite{Meaburn92,Meaburn98,Huggins92}. It was
anticipated both by Gibson~\cite{Gibson96} and
Schild~\cite{Schild96} that galactic dark matter is composed from
such planets. Each PGC contains about a billion of them.

\section{Galaxies} We may relate galaxies to the Jeans mechanism
at the end of the plasma epoch. The sole relevant aspect is then
the decrease of the speed of sound from plasma values to hot gas
values. Taking the geometric mean velocity $\overline V_S =
(V_S^{\rm plasma} V_S^{\rm gas})^{1/2}$= 874 km/s, we get the
Jeans scale $L_J=1.5$ kpc and corresponding mass

 \BEQ M_{gal}=\frac{\pi}{6}\rho_{B}L_J^3= \frac{\pi
{\overline V}_S^3}{6G^{3/2}\rho_B^{1/2}}=
1.4\cdot10^{11}M_\odot.\EEQ

\ni The corresponding CMB angle is $\theta_G=4.7'$ and the angular
index is $\ell_G=2300$. For this mass regime the formation time is
limited, because the sound speed continues to decrease to gas
values, from where on PGCs are formed. This explains why a lot of
baryons are not locked up in galaxies with their baryonic dark
matter, but located in intracluster and intercluster X-ray gas.
That gas has become hot, with temperatures up to 100 keV (1
keV$/k_B=1.16\cdot 10^7\K$), due to virialization after neutrino
condensation on clusters at $z\sim28$ or $t_{\nu{\rm c}}=120$
Myr~\cite{Nneutrino09}. At such high temperatures the gas may
allow nuclear fusion up to tellurium~\cite{Schatz2001}.


\section{Role of PGCs}
In some of them, still warm, collision processes have quickly led
to star formation, basically without a dark period, thus
transforming them into OGCs. Other PGCs transformed in ordinary
stars. In the major part of the PGCs the planets have frozen and
still persist without stars. These PGCs are in virial equilibrium
and act as ideal gas particles that constitute the galactic dark
matter. Their physical presence explains why the isothermal model
describes the basic features of galactic rotation curves so well,
that is, linear growth at small radius, plateau at large radii. To
improve the fit, one may consider mixtures with several isothermal
components~\cite{Nneutrino09}.

In the centers of galaxies near passings of PCGs will heat their
planets and induce star formation. Since this is mainly a
two-particle effect, the luminosity of a galaxy is expected to
relate to the PGC mass density as $L\sim \int\d^3r\,\rho_{PCG}^2$.
In the isothermal model $\rho_{PGC}=\sigma_v^2/2\pi Gr^2$, where
$\sigma_v$ is the velocity dispersion, so the $1/r^4$ fall off of
the integrand makes the luminosity finite. This results in the
scaling $L\sim \sigma_v^4/R$, which is the Faber-Jackson relation
~\cite{BinneyTremaine}, with an additional characteristic bulge
scale $R$. The Tully-Fisher relation is likewise explained, as it
involves the rotation velocity, which scales with $\sigma_v$.

In several instances the matrix of dark PGCs is revealed by new
star formation. When agitated by tidal forces the collision
frequency of the planets will increase causing re-evaporation of
the frozen gases, increased size and friction, and the possibility
of planet mergers to produce larger planets and eventually new
stars. The existence of galaxy dark matter in the form of clumps
of frozen primordial planets is clearly revealed in photographs of
galaxy mergers such as Tadpole, Mice and Antennae. On the
respective photographs one can see numerous bright clusters of
comparable size, that are identified here as PGCs turned into
young globular clusters. They are located in star wakes as the
merging galaxies enter each others dark matter halos and heat up
the planets in the PGCs on their path through the dark matrix. The
effect exists only within a certain radius, the boundary of the
PGC cloud.


\section{Role of H-He planets}
From the GHD scenario following decoupling, the first stars form
gently by a frictional binary accretion of still warm PFPs to form
larger planet pairs and finally small stars as observed in OGCs.
Thereby they create an Oort cavity as clearly exposed in e. g.
the Helix planetary nebula. Slow turbulent mixing from the rain of
planets will not mix away the dense carbon core.
By conservation of angular momentum
the star spins rapidly as it compresses producing strong spin
radiation along the spin axis and at the star equator. Stars can
form as binaries in PGC clumps.

At decoupling the entire baryonic universe turns to a fog of H-He
planet mass clouds. Due to the expansion they cool and the
freezing temperature of hydrogen and helium occurs at redshift $z
\approx 30$ a few hundred million years later, producing the
frozen dark baryonic planets in clumps predicted as the galaxy
dark matter ~\cite{Gibson96}. Neutrino condensation at $z\sim 28$
coincides with this, showing that the extra-galactic, gaseous H-He
planets heat up to become hot X-ray gas. That the dark matter of
galaxies should be planets of earth mass was independently
proposed by the Schild 1996 interpretation of 5 hr twinkling
periods in quasar microlensing observations
\cite{Schild96,Schild06}. Thousands of these planet crossings have
been observed by now.

As the universe expands the planets that did not turn into stars
freeze and the PGCs become less collisional and diffuse out of the
$3$ kpc scale protogalaxies to form the observed typical $R=100$ kpc dark
matter halos with a basically isothermal distribution. Consistency
of this picture is shown by the isothermal estimate $M_{\rm
gal}=2\sigma_v^2R/G\sim1.9\cdot 10^{12}M_\odot$ for $\sigma_v=200$
km/s.

Spiral galaxies reflect the accretion disks of dark
halo PGCs frictionally spiraling back toward the $L_N$ scale
protogalaxy remnant of ancient stars~\cite{Gibson08}. Violent
massive stars form at $L_{ST} = \varepsilon^{1/2} / ({\rho
G})^{3/4}$ as turbulent maelstroms at protogalaxy cores giving
spin radiation quasar and gamma ray burst events. For $\varepsilon
\sim 10^{-10}$ m$^2$/s$^3$ one derives a billion solar mass quasar
core.

When frozen, the H-He planets are too small to dim light, even
from remote sources, but they can account for dimming when they
are heated. Warm atmosphere diameters are $\approx 10^{13}$ m, the
size of the solar system out to Pluto, bringing them out of the
dark. The separation distance between planets is $\approx 10^{14}$
m , as expected if the PGC density of planets is the primordial
density $\rho_0 = 2 \times 10^{-17}$ kg m$^{-3}$. In planetary
nebula such as such as the nearby Helix, dark planets at the
boundary of the $3 \times 10^{15}$ m Oort cavity are evaporated.
HST optical images of Helix show $\approx 10^4$  ``cometary
knots'', planet-atmospheres $\approx 10^{13}$ m  which we identify
as H-He planets with metallicity, and Spitzer shows $\ge 10^5$ in
the infrared from the $10^3 M_{\odot}$ available~\cite{GS09}.
Meaburn et al. determine for cometary knots in Helix a mass of
$1.\, 10^{-5} M_\odot$, and conclude that the globules and tails
are dusty~\cite{Meaburn92}, and later report a mass of $2.\,
10^{-5} M_\odot$~\cite{Meaburn98}. Huggins et al ~\cite{Huggins92}
measure a mass of $5.\, 10^{-6}M_\odot$ via the radio measurement
of CO emission. These findings support the GHD prediction of earth
mass planets as the baryonic dark matter repository.

Dimming by dense $10^{13}$ m planet atmosphere gases or realistic
quantities of dust is negligible at the $20\%$ levels observed
~\cite{Aguirre 1999}.  Such large dimming of obscured lines of
sight observed in a planetary nebula (e.g. Helix) requires post
turbulent electron density forward scattering~\cite{GKB07} as
observed by radio telescope pulsar scintillation spectra, embedded
in the Kolmogorovian ``great power law on the sky''
~\cite{Gibson91}, which can now be understood from GHD as remnants
of supernova powered turbulent mixing in our local PGC
~\cite{GKB07}. The planet atmosphere  cross section for SNe Ia
dimming $\sigma \approx 10^{26}$ m$^{2}$ gives a photon mean free
path $ 1/n\sigma \approx 3 \times 10^{15}$ m from the primordial
PFP number density $n_0$, comparable to the observed Helix
planetary nebula shell thickness and consistent with $\approx 5\%$
of the SNe Ia lines of sight unobscured.
Clouds of warm H-He planets can be responsible for the
Ly$_\alpha$ forest -- not hydrogen clouds that have remained
undetected.

\section{Conclusion}  Within gravitational hydrodynamics (GHD)
for a flat space Friedman cosmology with (hot) neutrino cluster
dark matter we have described various structures that form due to
hydrodynamic instabilities in the baryonic plasma. The approach
accounts for a wealth of observations and relations between them.

The first step, when photon scattering from free electrons makes
the plasma too viscous, is the creation of proto-voids at
$z_\vf=5120$, before decoupling, that expand to present average
cosmic voids. Proto-supervoids occur too, but are rare. The matter
condenses in proto-galaxy-clusters while large scale turbulence at
boundaries of expanding voids in the plasma respect the earlier
determined  axis of evil.
The coincidence of turbulence properties in the CMB and turbulent
fluids supports the GHD model. Pancaked galaxy (cluster) structures
in between large voids arise dynamically, since the voids expand faster,
while the matter remains gravitationally bound.

The assumption that the kinetic energy of the protons, He atoms
and electrons is redistributed to the photons, explains the
micro-Kelvin scale of CMB fluctuations and the connection cold
spots -- voids, hot spots -- (super)clusters.

Before the viscous instability the plasma was quite homogeneous,
due to the free streaming neutrinos that damped inhomogeneities at
the free streaming scale of 10 kpc at $z_\vf$. So, irrespective of
their location on the sky, the fluctuations caused by the viscous
instability have the same order of magnitude. This well known
large angle correlation is in GHD an effect of simultaneity (of
void formation in the  homogeneous plasma), not of causality.

At decoupling first galaxies are formed and later primordial
globular clusters (PGCs) of about 38,000 solar masses. Some of
them turn into old globular clusters (OGCs) and others form the
stars in galaxy bulges. But most PGCs, millions per galaxy, remain
dark and constitute the galactic dark matter as an ideal gas,
which explains why the isothermal model describes galactic
rotation curves well. A galactic dark PGC matrix also accounts for
numerous young globular clusters seen in galaxy mergers.
Near PGC crossings in the center of galaxies will form stars,
which explains the Tully-Fisher and Faber-Jackson relations.

At the decoupling the viscous scale becomes much smaller which
turns all matter into H-He planets of a few earth masses. Some of
them coalesce to form the first small stars, but most freeze to
earth scale. The MACHO~\cite{Alcock98} and EROS ~\cite{Tisserand}
collaborations have searched in vain for such objects. Still, they
are not excluded because they do not occur uniformly but in PGC
clumps.  Theoretical descriptions of clumped MACHOs in the dark
halo were started in ~\cite{Holopainen}. But thousands of planets
have been observed in microlensing~\cite{Schild96,Schild06} and,
reheated, thousands more in planetary nebula such as Helix
~\cite{Meaburn92,Meaburn98,Huggins92,GibsonSchild07014}.
Hot H-He planet atmospheres may dim distant supernovas.

We have adopted one set of cosmological parameters, which
performed rather well, but not attempted an optimization. While in
the CDM model the main cause of clustering is dark matter with
baryons a second order effect, the GHD scales will be rather
sensitive to the precise parameter values. Large scale numerical
hydrodynamics simulations of separate steps of the fragmentation
process are expected to result in precise fits for the mass
fractions of baryons and neutrinos, and the Hubble constant.

We have not considered the large scale power spectrum~
\cite{WeinbergCosmo}. We may recall that there are reasons to
question whether baryons trace the neutrino dark matter
well~\cite{Nneutrino09}.

Let us finally see how some problems of the \LCDM paradigm
mentioned in the introduction are solved naturally in GHD.
Population III stars may have been rare, since reionization may
find its origin in neutrino condensation on galaxy
clusters~\cite{Nneutrino09}. Dwarf galaxies with a lot of
(baryonic!) dark matter may pertain to PGCs with incomplete star
formation. The related fact that OGCs often exhibit stars formed
at several epochs is likewise explained by further sets of
reheated, pre-existing H-He planets. Correlations in galaxy
structures are expected since they all form early; baryon acoustic
oscillations do not show up in GHD. Galaxy formation when the
universe was 4 -- 5 billion years young may refer to
proto-galaxies with late-stage star formation by close PGC
encounters. Dwarf satellites that swarm our galaxy just like its
stars may just relate to single PGCs with modest star formation,
popping out of the matrix of dark PGCs.

\end{document}